\documentclass[preprint,showpacs,preprintnumbers,amsmath,amssymb,prb,aps]{revtex4}

\usepackage{graphicx}
\usepackage{dcolumn}
\usepackage{bm}

\begin{document}

\bibliographystyle{prsty}

\title{
%
Electromagnetic radiation produced by avalanches in the
magnetization reversal of Mn$_{12}$-Acetate \vspace{-1mm}}

\author{
J. Tejada$^{1}$, E. M. Chudnovsky$^{1,2}$, J. M. Hernandez$^1$,
and R. Amigo$^1$}
\affiliation{\mbox{$^1$Department de F{\'\i}sica Fonamental, Universitat de Barcelona,}\\
\mbox{Diagonal 647, 08024 Barcelona, Spain}\\
\mbox{$^2$Department of Physics and Astronomy, Lehman College,
City University of New York,} \\ \mbox{250 Bedford Park Boulevard
West, Bronx, New York 10468-1589, U.S.A } }

\begin{abstract}
Electromagnetic radiation produced by avalanches in the
magnetization reversal of Mn$_{\bf 12}$-Acetate has been measured.
Short bursts of radiation have been detected, with intensity
significantly exceeding the intensity of the black-body radiation
from the sample.
The model based upon superradiance from inversely populated spin
levels has been suggested.
\end{abstract}


\maketitle


Populations of spin levels in molecular nanomagnets can be easily
manipulated by the magnetic field. For systems studied to date the
distances between the levels may go up to $0.5\,$THz. Sources and
detectors in this frequency range are scarce. Meantime the
electromagnetic radiation in the range $0.1 - 0.5\,$THz is
considered to be best for detecting small tumors and for security
screening at the airports \cite{Mittleman}. In this Letter we
report experimental studies of the electromagnetic radiation from
Mn$_{\bf 12}$-Acetate, performed with the help of an InSb
bolometer inside a SQUID magnetometer. Magnetic clusters
Mn$_{12}$-Ac have spin $S=10$ and high magnetic anisotropy that
corresponds to a $66\,$K energy barrier between spin-up and
spin-down states in zero field. The clusters form a centered
tetragonal crystal lattice inside which they weakly interact
through magnetic dipolar forces. At low temperature, due to the
high anisotropy, a crystal of Mn$_{12}$-Ac, can be prepared in a
metastable spin state \cite{Sessoli}. The crystals exhibit
extraordinary staircase magnetic hysteresis \cite{Friedman}
explained by resonant quantum spin tunneling. It also has been
known for some time \cite{Paulsen} that at low temperature and
high field-sweep rate, sufficiently large crystals exhibit an
abrupt reversal of the magnetic moment. This effect has been
attributed to a thermal avalanche in which the initial relaxation
of the magnetization towards the direction of the field results in
the release of heat that further accelerates relaxation. Direct
measurements of the heat emission by Mn$_{12}$-Ac crystals
\cite{Fominaya}, as well measurements of the magnetic relaxation
in pulse fields \cite{Barco}, supported this conjecture. The
avalanches have long been considered as undesirable events that
prevent experimentalists from studying spin tunneling in large
crystals. Our interest to avalanches was motivated by the fact
that they create a large inverse population of spin energy levels.

The experimental setup is schematically presented in Figure 1. The
sample was prepared by assembling together 20 single crystals of
Mn$_{12}$-Ac of total volume ~$V \approx 22\,$mm$^3$. The
conventional composition, lattice structure, and magnetic
properties of the crystals have been confirmed by chemical,
infrared, X-ray diffraction methods, and by SQUID magnetometry.
The crystals were glued together with their c-axes parallel to
each other within a maximum five-degree misalignment. The assembly
was placed inside a cylindrical waveguide, $5\,$mm in diameter,
with the c-axis approximately parallel to the axis of the guide.
The InSb bolometer was positioned inside the waveguide, $13\,$cm
above the sample located at the center of the measuring area of
the SQUID magnetometer. The cooling agent was helium gas
maintained at $1.8\,$K at the locations of the sample and the
bolometer. Before conducting measurements of the sample, we placed
the bolometer inside the measuring area of the magnetometer and
obtained the dependence of the resistance of the bolometer on
temperature and the magnetic field. The voltage drop on the
bolometer measured at a constant current goes down when it absorbs
radiation. The spectral bandwidth of the commercial InSb bolometer
is $60\,$GHz - $3\,$THz, and the response time when illuminated is
less than $1 \, \mu$s in the kelvin temperature range. To
calibrate the bolometer response to the radiation, we used a
heater that delivered measured heat pulses to the sample. The
temperature of the sample was monitored by a thermometer in a
thermal contact with the sample.

Figure 2A shows the magnetization curve obtained by sweeping the
magnetic field between $-2\,$T and $2\,$T at $1.8\,$K. For the
sweep rate between $5\,$mT/s and $27\,$mT/s the avalanches
reproducibly occur at ~$B \approx \pm 1.4\,$T. At the time of the
avalanche, the temperature of the crystals jumped by about $2\,$K,
from $1.8\,$K to $3.8\,$K. Figure 2B shows two bursts of the
radiation measured by the InSb bolometer during the total
magnetization cycle at the field-sweep rate of $27\,$mT/s. These
bursts have been observed in hundreds of hysteresis cycles and are
completely reproducible in terms of their position and height.
They coincide precisely with the occurrence of the avalanches. The
time dependence of the burst is blown in Fig. 2C. Each burst is
characterized by a steep decrease of the voltage drop on the
bolometer within a time of the order of $0.03\,$s. It is followed
by the slow increase of the voltage towards the noise level, which
we interpreted as a thermal equilibration of the system. A similar
equilibration has been observed after supplying a heat pulse to
the sample. However, the short radiation burst preceding the
equilibration appears to be a unique signature of the
magnetization avalanche. The total magnetic energy released in the
avalanche was $B{\Delta}M \, \sim \, 2\,$mJ. Experimenting with
heat pulses, we found that four times this energy (resulting in
the $12\,$K temperature of the sample) was required to produce the
bolometer signal of the level of the radiation burst from the
avalanche.

The spin Hamiltonian of Mn$_{12}$-Ac molecular cluster in the
magnetic field applied along the c-axis, is
\begin{equation}\label{Hamiltonian}
{\cal{H}}=-DS_{z}^{2}-FS_z^4-bS_{z}+{\cal{H}}'\,,
\end{equation}
where ~${\bf S}$~ is the spin of the cluster, ~$D=0.55\,$K~ and
~$F=1.1{\times}10^{-3}\,$K~ are uniaxial anisotropy constants,
~$b=g{\mu}_{B}B$~ is the reduced magnetic field (with ~$B$~ being
the actual field, ~$g=1.94$~ being the gyromagnetic factor, and
~${\mu}_{B}$~ being the Bohr magneton), and ~${\cal{H}}'$~
contains smaller terms that do not commute with ~$S_{z}$~ (see for
details Ref. \cite{Hill-Kent}). The energy levels of
(\ref{Hamiltonian}) are given by
\begin{equation}\label{levels}
E_m=-Dm^2-Fm^4-bm \,,
\end{equation}
with $m$ being the magnetic quantum number for the spin of the
molecule ${\bf S}$. At ~$b=kD$, with ~$k=0, \pm 1, \pm 2, ...$,
the levels ~$m$~ and ~$m'$~ satisfying $m+m'=-k$~ are nearly
degenerate, as is illustrated in Figure 3 for ~$k=3$. This
resonance approximately corresponds to the field at which the
avalanches have been observed.

For an individual molecule the rate of the spontaneous decay from
the level ~$m$~ to the level ~$m+1$~ with the emission of light is
\begin{equation}\label{em1}
{\Gamma}_{1}=\frac{2g^{2}{\mu}_{B}^{2}}{3{\hbar}^{4}c^{3}}(S+m)(S-m+1)(E_{m}-E_{m+1})^{3}\;.
\end{equation}
With the help of Eq.\ (\ref{Hamiltonian}) one finds that at
~$k=3$~ the frequencies of the corresponding radiation are between
$f \approx 80\,$GHz for the ~$m=1 \, \rightarrow \, m=2$~
transition and $f \approx 350\,$GHz for the ~$m=9 \, \rightarrow
\, m=10$~ transition, that is, in the range of the InSb bolometer.
The corresponding wavelengths are between $3.8\,$mm for the ~$m=1
\, \rightarrow \, m=2$~ transition and $0.85\,$mm for the ~$m=9 \,
\rightarrow \, m=10$~ transition. At ~$k=3$~ and ~$m=1$~ the rate
~${\Gamma}_{1}$~ of Eq.\ (\ref{em1}) is of the order of
$10^{-7}\,$s$^{-1}$, that is, the lifetime of the corresponding
excited state with respect to the photon emission is of the order
of one year. Other individual magnetic dipole transitions in
Mn$_{12}$-Ac have similarly low rates. However, as first noticed
by Dicke \cite{Dicke}, $N_m$ dipoles confined within a volume of
size ~$d$~ which is small compared to the wavelength of the
radiation ~$\lambda$, cannot radiate independently. At ~$d <
{\lambda}$~ a spontaneous phase locking of the atomic dipoles
occurs that results in the coherent radiation burst with the power
~$P_{SR} \propto N_m^2$~ emitted within a time of the order of
~${\tau}_{SR} \, {\sim} \, 1/N_{m}{\Gamma}_m$. This phenomenon,
known as the superradiance or superfluorescence, has been widely
observed in gases on increasing the concentration of the gas
confined within a volume of dimensions ~$d < {\lambda}$ (see
reviews \cite{Haroche-Men'shikov}). In our case the condition ~$d
< {\lambda}$~ is fulfilled at least for the first few transitions
starting from ~$m=1$. For ~$m=1$ the superradiance rate would then
be ${\Gamma}_{SR} = 1/{\tau}_{SR} = N_{1}{\Gamma}_{1}$, where
~$N_1 \, {\sim} \, N\exp(-U_{eff}/T)$~ is the population of the
$m=1$ level, $N$ is the total number of Mn$_{12}$-Ac molecules
available for the relaxation and ~$U_{eff}$~ is the effective
energy barrier between the two wells shown in Fig. 3. For our
sample, ~$N \, {\sim} \, 6{\times}10^{18}$ and ~$U_{eff} \approx
47\,$K. The latter corresponds to the distance from $m=-10$ to the
$m=-4,1$ resonance that dominates the magnetic relaxation at ~$k
\approx 3$ and ~$T=3.8\,$K. Correspondingly, ~$\exp(-U_{eff}/T) \,
{\sim} \, 4{\times}10^{-6}$ and $N_{1} \, {\sim} \,
2{\times}10^{13}$. The rate of the electromagnetic decay of the
$m=1$ state then increases from ~${\Gamma}_{1} \, {\sim} \,
10^{-7}\,$s$^{-1}$~ to ~${\Gamma}_{SR}=N_1{\Gamma}_1 \, {\sim} \,
2{\times}10^6\,$s$^{-1}$. The total duration of the relaxation due
to the continuous supply of the population of the $m=1$ level is
given by ~${\tau} = {\tau}_{SR}\exp(U_{eff}/T) \, {\sim} \,
0.1\,$s, which is in a remarkable agreement with experiment.

In our model the relaxation from the upper levels in the right
well in Fig. 3 is dominated by the emission of photons, while at
the lower levels it must switch to the emission of phonons due to
the violation of the condition ~$d < {\lambda}$~ at the bottom of
the well. This condition is one of the two basic requirements for
the superrradiance to occur. The other condition is that
~${\Gamma}_{SR}$~ exceeds the rate of decoherence of the phases of
the magnetic dipoles. In our case this decoherence is provided by
spin-phonon transitions at a rate ${\Gamma}_{phon} \, < \,
10^6\,$s$^{-1}$ \cite{GC}. Remarkably, the superradiance rate
computed above, ${\Gamma}_{SR}\, {\sim} \,
2{\times}10^6\,$s$^{-1}$, indeed, exceeds the phonon rate. Given
the fact that our model also provides the correct duration of the
avalanche, this is hardly a coincidence. It is, therefore,
conceivable that the magnetization avalanche is an electromagnetic
phenomenon to the same degree as it is a phonon runaway. This
conjecture is supported by the experimental fact that avalanches
only occur when the total number of molecules in a crystal
assembly exceeds a critical value, $N \, {\sim} \, 10^{18}$,
needed to ignite superradiance.

The work at the University of Barcelona has been supported by the
European Commission through Contract No. IST-2001-33186 and by the
Spanish Government through Contract No. MAT-2002-03144. The work
of E.M.C. has been supported by the U.S. National Science
Foundation through Grant No. EIA-0310517.



\newpage

\begin{figure}[t]
\centering
\includegraphics[width=12cm]{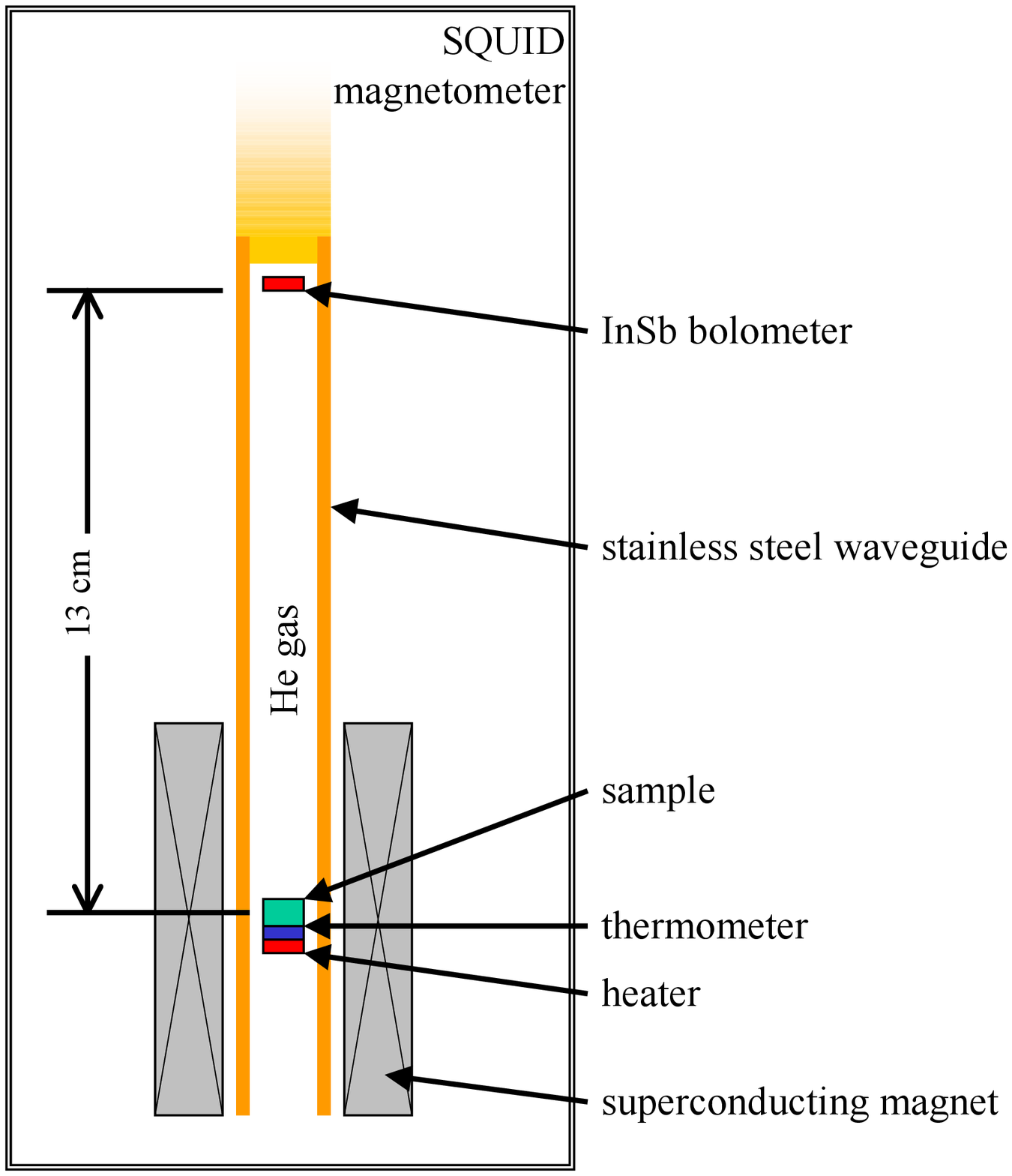}
%
\caption{ \label{fig_setup} Experimental setup.}
\end{figure}
\newpage
\begin{figure}[t]
\centering
\includegraphics*[width=10cm]{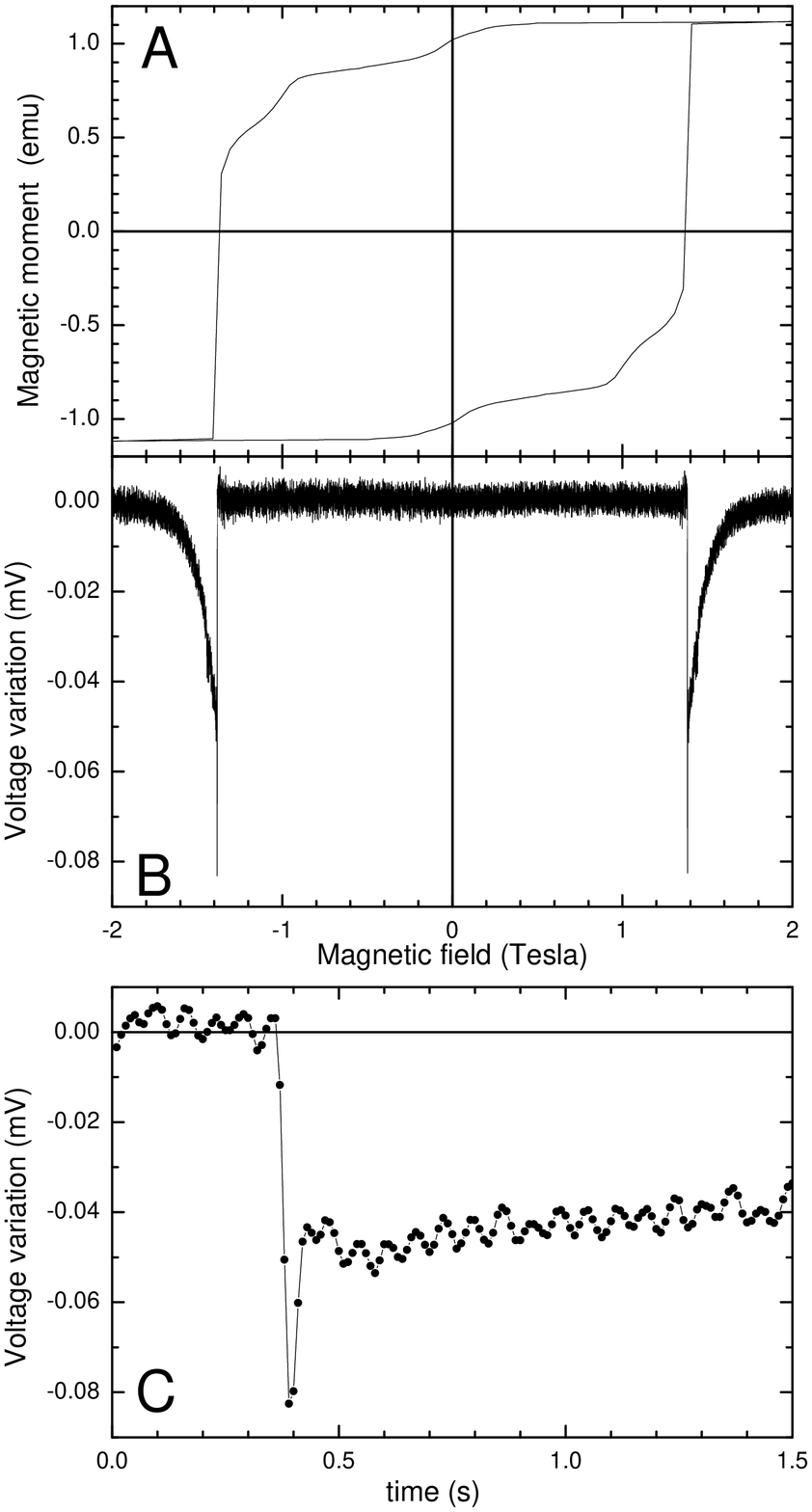}
\caption{ \label{fig_radiation} (A) Hysteresis curve of the
20-crystal sample of Mn$_{12}$-Ac at $T=1.8\,$K and the
field-sweep rate of $27\,$mT/s. Avalanches in the magnetization
reversal reproducibly occur at ~$B \approx \pm 1.4\,$T; (B)
Variation of the voltage drop on the InSb bolometer as a function
of the magnetic field. The radiation bursts detected by the
bolometer coincide with the magnetization reversal; (C) Time
dependence of the bolometer voltage for one of the radiation
bursts seen in Fig. 2B. }
\end{figure}
\newpage
\begin{figure}[t]
\centering
\includegraphics[width=10.0cm]{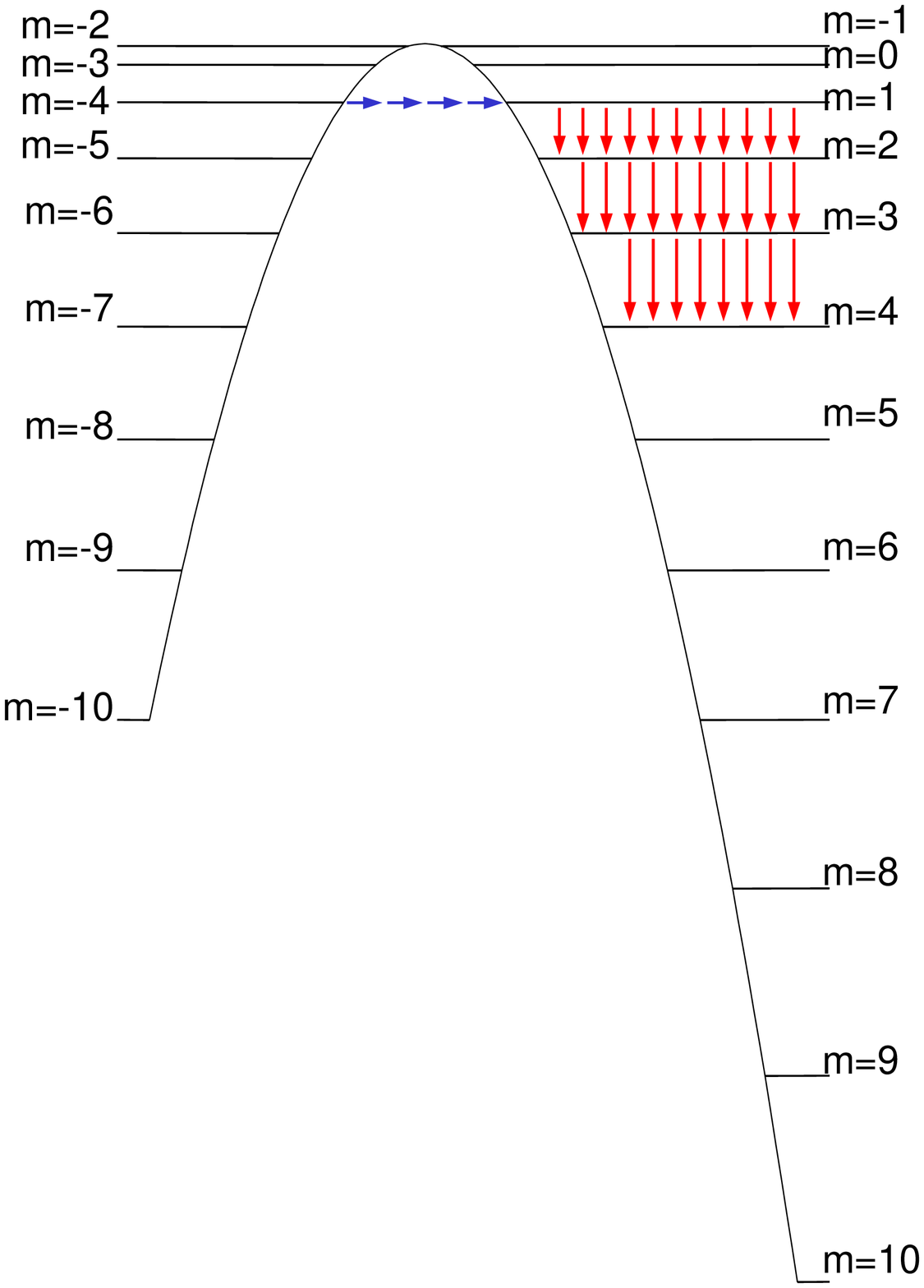}
%
\caption{Spin levels of Mn$_{12}$-Ac in the magnetic field
corresponding to the third resonance, ~$k=3$. The horizontal axis
represents the magnetic quantum number ~$m$. Arrows show the
relaxation path from the initial state magnetized in the negative
z-direction to the final state magnetized in the positive
z-direction.}
\end{figure}

\end{document}